\documentclass[aps,prd,twocolumn,showpacs,amssymb]{revtex4-1}
\usepackage{graphicx}
\usepackage{bm}
\usepackage{amsmath}
\usepackage{amssymb}%
\usepackage{float}
\newcommand{\jcap}{Journal of Cosmology and Astroparticle Physics}  
\newcommand{\Apjl}{The Astrophysical Journal Letters}  
\newcommand{\aap}{Astronomy and Astrophysics}  
\newcommand{\mnras}{Monthly Notices of the Royal Astronomical Society}
\newcommand{\araa}{Annual review of astronomy and astrophysics}
\newcommand{\apss}{Astrophysics and Space Science}  
\newcommand{\Apj}{The Astrophysical Journal}  

\begin{document}
\title{Analytical treatment for the development of electromagnetic cascades in intense magnetic fields}

\author{Jie-Shuang Wang$^{1,2}$}\email{jiesh.wang@gmail.com}
\author{Ruo-Yu Liu$^{2,3}$}\email{ruoyu.liu@desy.de}
\author{Felix Aharonian$^{2,4}$}\email{Felix.Aharonian@mpi-hd.mpg.de}
\author{Zi-Gao Dai$^{1,5}$}\email{dzg@nju.edu.cn}
\affiliation{$^1$ School of Astronomy and Space Science, Nanjing University, Nanjing 210093,China\\
$^2$ Max-Planck-Institut f\"ur Kernphysik, Saupfercheckweg 1, D-69117 Heidelberg, Germany\\
$^3$ Deutsches Elektronen Synchrotron (DESY), Platanenallee 6, D-15738 Zeuthen, Germany\\
$^4$ School of Cosmic Physics, Dublin Institute for Advanced Studies, 31 
Fitzwilliam Place, Dublin 2, Ireland\\
$^5$ Key Laboratory of Modern Astronomy and Astrophysics, Nanjing
University, Ministry of Education, Nanjing 210093, China}

\newcommand{\be}{\begin{equation}}
\newcommand{\ee}{\end{equation}}
\def\ba{\begin{eqnarray}}
\def\ea{\end{eqnarray}}
\def\bc{B_{\rm cr}}
\def\c0{\chi_{0}}
\def\cp{\chi}
\def\cg{\chi_{\gamma}}
\def\ce{\chi_{e}}
\def\cc{\chi_{\rm c}}
\def\cb{\chi_{\rm b}}
\def\bp{B_{\perp}}
\def\cm{\chi_{\rm m}}
\def\um{\Upsilon_{\rm c}}
\def\ug{\Upsilon_{\gamma}}
\def\ft{f_{\rm si}}
\begin{abstract}
In a strong magnetic field, a high-energy photon can be absorbed and then produce an electron-positron pair. The produced electron/positron will in turn radiate a high-energy photon via synchrotron radiation, which then initiates a cascade. We built a one-dimensional Monte Carlo code to study the development of the cascade especially after it reaches the saturated status, when almost all the energy of the primary particles transfers to the photons. 
The photon spectrum in this status has a cutoff due to the absorption by magnetic 
fields, which is much sharper than the exponential one. 
Below the cutoff, the spectral energy distribution (SED) manifest itself as a 
broken power-law with a spectral index of $0.5$ and $0.125$, respectively, below and above the broken energy. The SED can be fitted by a simple analytical 
function, which is solely determined by the product of the cascade scale $R$ and the magnetic field perpendicular to the motion of the particle $\bp$, with an accuracy better than 96\%. The similarity of the spectrum to that from the 
cascade in an isotropic black-body photon field is also studied.
\end{abstract}
\maketitle
\section{Introduction}
The process of an electromagnetic (EM) cascade in a magnetic field is of important relevance in high-energy astrophysics, such as the detection of ultra-high-energy photons at the Earth \citep{McBreen1981,*Aharonian1991,*Plyasheshnikov2002}, and the nonthermal emission in some extreme astrophysical objects with intense magnetic fields. The EM cascade in the magnetic field requires an extreme condition, in terms of $\cp=\epsilon \bp/m_e c^2 \bc \gtrsim 1$, where $\bc=4.414\times 10^{13}\,$G is the quantum critical magnetic field, $\bp$ is the magnetic field perpendicular to the momentum of the electron or the photon, $\epsilon$ is the energy of the incident electron or photon, $m_e$ is the electron mass and $c$ is the speed of light. More specifically, in the condition of $\cp\gtrsim1$, the one-photon pair production process can happen, in which the magnetic field absorbs a photon with energy larger than $2m_ec^2$ and produces an $e^\pm$ pair \citep{Erber1966,Daugherty1983,Baring1988, Akhiezer1994,Anguelov1999}. In addition, the synchrotron radiation by an electron (or positron, hereafter we do not distinguish the positron from the electron) needs to be modified in the quantum electrodynamics (QED) regime. In the classical regime, the typical synchrotron photon energy is $\hbar \omega_a=0.29 \hbar \omega_c\simeq 0.44\cp\gamma m_e c^2$, where $\omega_c\simeq 1.5\gamma^2e\bp/m_ec$. The classical synchrotron radiation is valid only when $\hbar \omega_a \ll\gamma m_e c^2$ is satisfied, namely $\cp\ll1$. In the QED regime, the average energy of the synchrotron photon can be comparable to the primary electron energy \citep{Erber1966,Baring1989,Aharonian2003}. Thus, the synchrotron radiation from the secondary pairs will regenerate high-energy photons which would again produce pairs, and thus an EM cascade will be triggered.

Various attempts have been made to study the EM cascade in the magnetic field, such as solving the kinetic equations directly \citep{Akhiezer1994}, using Monte Carlo simulations \citep{Anguelov1999}, and solving the adjoint cascade equations numerically \citep{Aharonian2003}. These researches studied the properties of cascades, but only in the first few hundreds of the propagation lengths of the primary particle, where the interaction probability is unity in one propagation length as defined in Eq.~(3). However, in reality, the cascade generally develops very deeply in the astrophysical scale, which is much larger than a few hundred propagation lengths of the primary particle. On the other hand, those numerical methods are sometimes quite expensive, although effective. In this work, we aim to constitute a simple yet accurate analytical solution for the development of the cascade in the case assuming that the spatial scale of the region where the cascade has developed is much larger than the mean propagation length of the electrons and gamma rays due to the synchrotron radiation and the pair production, respectively.

Such cascades could happen in some astrophysical sources and be responsible for their high-energy gamma-ray ($\gtrsim$GeV) radiation. Based on the Hillas criterion \citep{Hillas1984}, the maximum energy of a particle with charge number $Z$ can be accelerated by an astrophysical object is $E_{\rm max}\sim 300 ZBR_a$ eV, where $R_a$ is the characteristic spatial scale of the acceleration region. Denoting the energy of the secondary photon or electron produced by the accelerated particle with $\alpha E$ (with $\alpha <1$), we have $\cp\sim1000\alpha Z(B/10^4{\rm G})^2(R_a/10^{12}{\rm cm})$. Usually, we expect $\alpha \sim 0.001$-$0.1$ from the Bethe-Heitler process or hadronic interactions of accelerated cosmic rays, so these astrophysical objects (summarized in \citep{Aharonian2002,*Ptitsyna2010}) can in principle meet the condition $\cp >0.1$, and hence be potentially interesting sources for such a process. 

The paper is organized as follows: we study the development of the synchrotron-pair cascade with a Monte Carlo method in Sec. II. In Sec. III, we fit the SED obtained by the Monte Carlo simulation with a simple function. In Sec. IV, we study the EM cascade in the presence of a black-body photon field, and discuss similarity between the cascade in a magnetic field and in a photon field with a black-body distribution. The summary is given in Sec. V.

\section{one-dimensional Monte Carlo simulation of photon-pair cascades in intense magnetic fields}
The main processes in the cascade are the synchrotron radiation and pair production. 
The differential probability (in units of cm$^{-1}$) of an electron with energy $E=\gamma m_ec^2$ producing a photon with energy $\epsilon$ via synchrotron radiation $(P_{\rm syn})$, and the probability of a photon with energy $\epsilon$ producing an electron with energy $E$ through pair production $(P_{\rm pp})$ are given by \citep{Akhiezer1994,Baring1988,Anguelov1999},
\ba
\nonumber
P_{\rm syn}(\ce,\cg)d\cg ={e^2 m_e \bp d\cg\over\sqrt{3}\pi\hbar^2\bc \ce^2}\times  ~~~~~~~~~~~~~\\ 
\left(\int_{y}^{\infty}K_{5/3}(x)dx+{\cg^2\over \ce(\ce-\cg)}K_{2/3}(y)\right),\label{eq:syn}
\ea
and
\ba
 \nonumber
P_{\rm pp}(\cg ,\ce)d\ce ={e^2 m_e\bp d\ce \over\sqrt{3}\pi\hbar^2 \bc \cg^2}\times~~~~~~~~~~~~~~~~\\
\left(\int_{-y}^{\infty}-K_{5/3}(x)dx+{\cg^2\over \ce(\cg-\ce)}K_{2/3}(-y)\right),\label{eq:pp}
\ea
respectively, where $\ce=E\bp/\bc m_ec^2$ with for an electron, $\cg=\epsilon\bp/\bc m_ec^2$ 
for a photon, and $y={2\cg/ 3\ce(\ce-\cg)}$. 
The modified Bessel functions $K_{5/3}$ and $K_{2/3}$ are used in these formulas. 

\begin{figure}[t]
\includegraphics[width=\columnwidth]{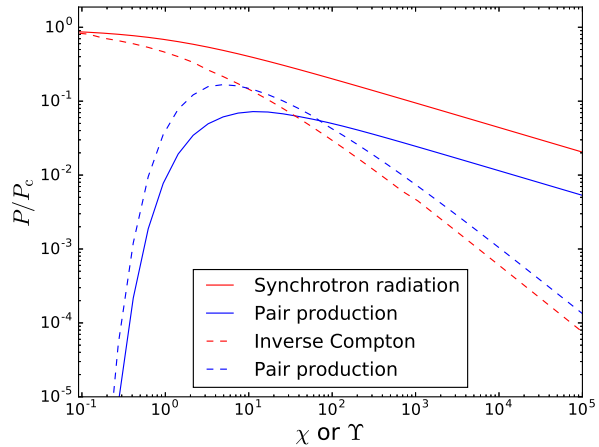}
\caption{\label{fig1} The solid curves are the normalized interaction 
probabilities ($P/P_{\rm c}$) of the synchrotron radiation (solid red curve) 
and pair production (solid blue curve) in the magnetic fields as functions of $\chi$. 
The dashed curves are the normalized interaction probabilities  
of the inverse Compton radiation (dashed red curve) and pair production 
(dashed blue curve) in the black body radiation fields as functions of $\Upsilon$.}
\end{figure}

The development of the cascade is characterized by the mean {\it propagation} length, which is related to the probability of  interaction as
\be
L=P^{-1}.
\ee
Note that in the case of the synchrotron radiation in the classical regime, when the electron in a single interaction with the magnetic field looses a very small fraction of its energy, $\Delta E << E$, this definition becomes rather meaningless. Therefore the radiation length is defined as $L=cE/(dE/dt)$ where $dE/dt$ is the continuous energy loss rate of electrons. In the quantum regime of the synchrotron radiation, when the electron loses a significant fraction of its energy, Eq.(3) appropriately describes the propagation length as it does for the pair production. In the deep QED regime ($\cp\gg1$), the probability of the synchrotron radiation and pair production have asymptotic forms:
$P_{\rm syn,tot}=6.23\times10^{-6}\bp\cp_e^{-1/3}$ cm$^{-1}$, 
and $P_{\rm pp,tot}=1.62\times10^{-6}\bp\cp_{\gamma}^{-1/3}
$ cm$^{-1}$, respectively \citep{Akhiezer1994,Anguelov1999}.
Thus the interaction probabilities of both processes are scaled as $\cp^{-1/3}$.
The probability of the classical synchrotron radiation ($\cp\ll1$),
can be obtained by $P_{\rm syn,tot}=P_{\rm c}=2 e^4 \bp^2\gamma^2/3 
m_e^2 c^4 \hbar \omega_a=6.56\times10^{-6}\bp$ cm$^{-1}$.
While the probability of the pair production at $\cp<1$ is $P_{\rm pp,tot}=
9.84\times10^{-7}\bp\exp{(-8/3\cp_{\gamma})}$ cm$^{-1}$ \citep{Erber1966}.
We show the normalized interaction probabilities $P/P_{\rm c}$ of 
the synchrotron radiation and pair production with solid curves in Fig. \ref{fig1}. 
Photon splitting, whose total probability is $P_{\rm sp,tot}=8.4\times10^{-15}\bp\cg^5$ 
cm$^{-1}$ \citep{Adler1970,*Adler1971,*Baring1991,*Harding1997,*Baring2001}, 
will dominate over the pair production, if 
$\cg<0.086$. However as will be shown below, in our consideration, only the photons with $\cg\gtrsim0.1$ 
will be absorbed by magnetic fields. Thus photon splitting is not important
in this study, and we will neglect it in the following calculation. 

We built a one-dimensional Monte Carlo code to study the synchrotron-pair cascade in the intense and homogeneous magnetic field. 
According to \citep{Baring2005} and \citep{Harding1987}, the high-energy secondary particle will generally follow the same direction of the primary particle in an arbitrarily orientated magnetic field, if the energy of emitted secondary particle is $\gg m_ec^2$, which is usually true in the considered scenario of this work. Thus, the one-dimension assumption works well for the high-energy range of the cascade. This is also the reason why the magnetic field is usually assumed to be perpendicular to the momentum of the particle in the simulations \citep[e.g.][]{Anguelov1999,Aharonian2003}. Also we find that the interaction probability of the primary particle and the spectrum of the secondary particle depend on $\bp$ only [e.g. see Eqs.~(\ref{eq:syn}) and (\ref{eq:pp})]. Therefore, no matter how is the magnetic field orientated with respect to the particle's velocity, we will only take the perpendicular component $\bp$.
The characteristic cascade scale in this paper is chosen to be much larger than the propagation length of the initial particle [$R\gg L(\chi_{0},\bp)$], so that the cascade is saturated, namely, the energy of primary particle goes eventually into the photons with energies below the pair production limit. We note that although one would expect that electrons could be deflected in the magnetic field so that the scenario may not be described in one dimension. However, for most cases, we concentrate on the photons with $\cg\gtrsim 10^{-5}$, which are produced by electrons with $\ce\gtrsim 5\times10^{-3}$. The Larmor radii of electrons above this energy are much larger than the energy loss length, so these photons are emitted before the electrons are significantly deflected. What's more, even if the motion of electrons are significantly deflected away from the initial direction by the magnetic field, the one-dimensional treatment still works as long as the pitch angle of the particle remains almost unchanged. In this case, $R$ is just not the displacement of the particle but the distance covered by the particle. Thus, a one-dimensional treatment is sufficient for this study. 

We then let particles propagate in the magnetic field with the step size $\Delta s$, which is from the condition $\Delta s\cdot P_{\rm tot}=0.05$. The energies of the secondary photons and electrons are determined by Eqs.~(\ref{eq:syn}) and ~(\ref{eq:pp}) respectively,  with 400 bins being divided from $\chi_{\rm max}/400$ to $\chi_{\rm max}$. Here $\chi_{\rm max}$ equals to the energy of the incident particle $\chi_0$ for the pair production and for the synchrotron radiation with $\chi_0>0.15$, while $\chi_{\rm max}$ equals to $3\hbar\omega_c\bp/B_{\rm cr}m_ec^2$ for the synchrotron radiation when $\chi_0<0.15$.
We trace each secondary particle in terms of its energy and distance propagated, regardless of the directions of their velocities. The energies of the particles will be recorded after certain expected cascade development scale $R$. To reduce as much as possible the statistical error in the Monte Carlo simulation and in the meantime keep the calculation not too expensive, we select the number of the primary particles at injection from hundreds to several thousands in different cases so that there will be at least $1000$ particles in each energy bin of the recorded spectrum of secondary particles. 

\begin{figure}[t]
\centering
\includegraphics[width=\columnwidth]{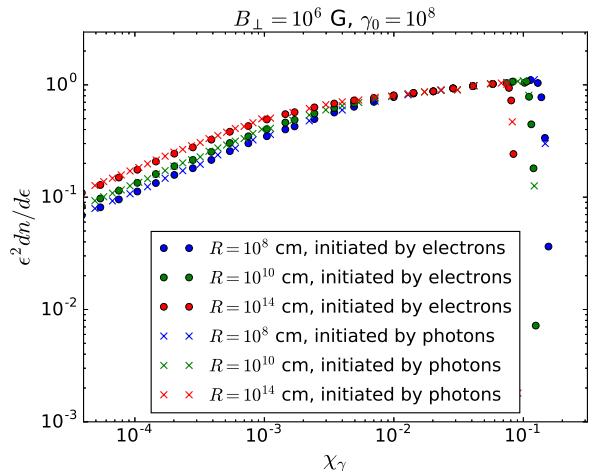}
\caption{The SEDs ($\epsilon^2 dn/ d\epsilon$) of the synchrotron-pair 
cascades initiated by electrons and photons are shown with filled circles and crosses, respectively. 
The energy spectra of cascades in scales of $R=10^8$ cm, $10^{10}$ cm, 
and $10^{14}$ cm are show with blue, green, and red, respectively.
\label{fig2} } \end{figure}

We firstly compare the cascades initiated by electrons and by photons. The background magnetic field is fixed at $10^6$~G in both two scenarios, and the energies of the primary electrons or photons are $2\times10^8 m_e c^2$. In Fig.~\ref{fig2}, we show the spectrum of photons in the cascades at three cascade scales, i.e., $R=10^8\,$cm, $10^{10}\,$cm and $10^{14}\,$cm respectively.We can see that the photon spectrum in the electron-initiated cascade and that in the photon-initiated cascade are identical to each other. Thus, the type of the primary particles makes no difference to cascade as long as the cascade is saturated. 

\begin{figure}[t]
\centering
\includegraphics[width=\columnwidth]{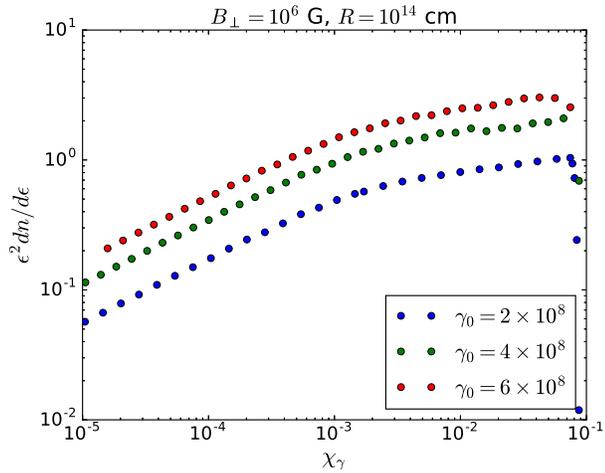}
\caption{The SEDs ($\epsilon^2 dn/ d\epsilon$) of the electron 
initiated synchrotron-pair cascades. The background magnetic field is $\bp=10^6$ G, and 
the cascade scale is $R=10^{14}$ cm. The Lorentz factor of primary particles are shown with different colors, as listed in the legend.
\label{fig3} }
\end{figure}

Secondly, we study the influence of the energy of the initial particle on the cascade, employing electron as the primary particle. Three different initial Lorentz factors of electron, say, $2\times 10^8$, $4\times10^8$ and $6\times10^8$, are considered, while we fix the background magnetic field at $\bp=10^6$ G, and the cascade scale at $R=10^{14}$~cm. The results are shown in Fig.~\ref{fig3}.  
It can be found that the amplitude of energy spectra is proportional to the energy of the primary 
particle which is obvious from the condition of energy conservation, while the spectral shapes in the three cases coincide with each other precisely. Thus, hereafter 
we will normalize the SED with $\epsilon^2 dn/\epsilon_0 d\epsilon$. 
The advantage of using this normalized SED is that it is only a 
function of $\cg$, i.e.,
\ba
\epsilon^2 dn/\epsilon_0 d\epsilon=f(\cg).\label{eq1}
\ea
Due to the conversation of energy $\int \epsilon dn/\epsilon_0\approx1$ for 
the saturated cascade, we have 
\be
\int f(\cg)d\cg/\cg\approx1.\label{eq2}
\ee

\begin{figure}[t]
\centering
\includegraphics[width=\columnwidth]{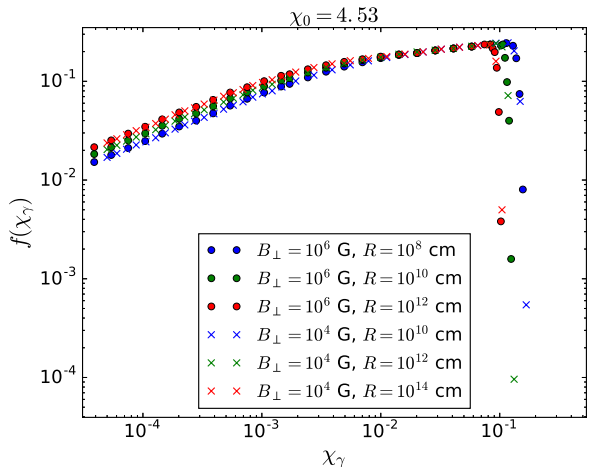}
\caption{Normalized SEDs ($f(\cg)=\epsilon^2 dn/\epsilon_0 d\epsilon$) of the 
synchrotron-pair cascades.in different magnetic fields: $10^6$ G (filled circles) and 
$10^4$ G (crosses). Spectra of different cascade scales are shown with different colors. 
$\c0=4.53$ is fixed for the initial particle.
\label{fig4} } \end{figure}

Finally, we study the cascades in magnetic fields of different strength: $10^6$\,G and 
$10^4$\,G. The energies of primary particle are chosen to make $\c0=4.53$ in both two cases which is also identical to that in Fig.~\ref{fig2}. The results are 
shown in Fig. \ref{fig4}. We find that for different background magnetic fields, the 
energy spectra (as functions of $\cg$) and their evolutions are identical, as long as the quantity $R\bp$ 
keeps the same. This can be understood through Eqs.~(\ref{eq:syn}) and 
(\ref{eq:pp}). Specifically, given the same $\chi_0$ for electrons, their propagation lengths 
are then inversely proportional to the magnetic fields, i.e., $L_{\rm syn} \propto 1/\bp$. 
Thus, if $R\bp$ is a constant for these electrons, their interaction 
probabilities will be exactly the same. The produced photon spectra are also identical 
since the spectral shape is determined only by $\ce$ as shown in Eq.~(\ref{eq:syn}). 
In brief, the normalized SED in the saturated cascade is only determined by the product $R\bp$, neither the 
type nor the energy of the primary particle.

\section{spectral fitting of the cascades in intense magnetic fields}

The SED of the photons in the cascade can be treated as consisting of three parts. 
From high energy to low energy, the first part is a sharp cutoff at $\chi\sim0.1$, which is due to the absorption of the highest energy photon by the magnetic fields. The cutoff energy $\cc$ could be found via equating the propagation length of the particle to the cascade scale, namely, $L(\cc)=R$. Empirically, we obtain $\cc=2.67/\left[2.30\log_{10} \left(R\bp/{\rm (cm\cdot G)}\right)-13.74\right]$ valid for $\log_{10} \left[R\bp/{\rm (cm\cdot G)}\right]\geq12$. The spectrum starts to drop roughly at $\sim0.95 \cc$. 

The second part is in the range of $\cb<\cg<\cc$ with a power-law 
index at $0.125$ approximately, where $\cb$ is the break energy separating the second part 
from the third part. Photons in this energy range is emitted by electrons produced by photons with energy $\cg>\cc$ via the one-photon pair production in the magnetic field. Thus, $\cb$ corresponds to the typical energy of the synchrotron photon emitted by electrons that are produced in the pair production of cutoff 
photons with energy $\cc$. Considering $\cc\sim0.1$, the energy of the generated 
electrons are around $\ce\sim 0.5 \cc$. Using the classical synchrotron radiation formulas, 
the break energy can be obtained by $\cb\approx0.38\cc^2$. The normalized spectrum between $\cb$ and $\cc$ is quite robust, and can be fitted by $f(\cg)= 0.323 \cg^{0.125}$, independent of the product $R\bp$.

As mentioned above, $\cb$ corresponds to the typical energy of the synchrotron radiation by the electrons of the lowest energy generated in the pair production. Thus, photons with energies $<\cb$, which constitute the third part of the spectrum, are emitted by those electrons cooled from higher energies via synchrotron radiation in the classical regime. Without injection at such energies, the electron spectrum is solely regulated by the synchrotron cooling, resulting in a spectrum $\propto \gamma^{-2}$. Hence, the third part of the SED behaves as $f(\cg)\propto \cg^{0.5}$. 

\begin{figure}[]
\centering
\includegraphics[width=\columnwidth]{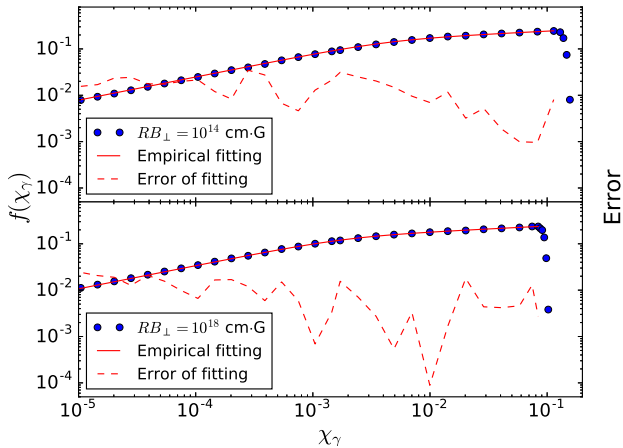}
\caption{The normalized SEDs for $R\bp=10^{14}$ cm$\cdot$G, and 
$10^{18}$ cm$\cdot$G are presented. The error is defined as 
$|$simulated value - analytical value$|$/simulated value.
\label{fig5} } \end{figure}

\begin{figure}[]
\centering
\includegraphics[width=\columnwidth]{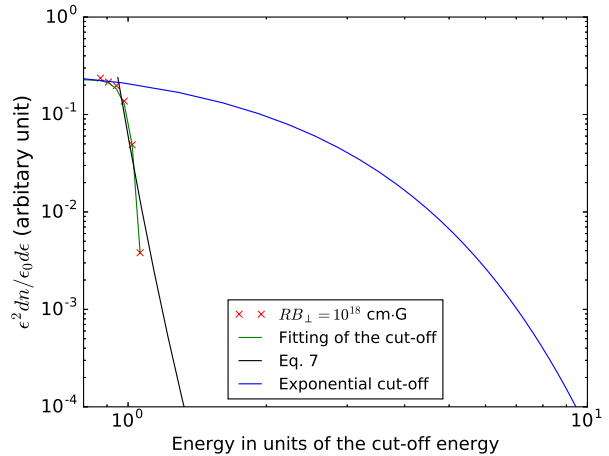}
\caption{The spectral cutoff from the cascade spectrum (red crosses), 
its analytical fitting (green solid line), the space-integrated spectrum by Eq.~(\ref{eq:totcut}) 
(black solid line). The $x$-axis is the photon energy in unit of the cutoff energy, i.e., 
$\chi/\chi_c$. For comparison, we also show the cutoff regime of spectrum of the classical synchrotron radiation of an electron with $\ce=3\times10^{-4}$ and $\bp=10^4$ G (blue solid line) which represents the exponential cutoff. The $x$-axis for this curve is $\omega/\omega_c$ with $\omega_c$ being the classical synchrotron radiation frequency.
\label{fig6} } \end{figure}

We find that we can use an empirical function $f(\cg)$ 
to fit the SED below the cutoff (namely $\cg<0.95\cc$) very well,
\be
f(\cp)=\left[{1\over 1/(a\cg^{0.5})^{4}+1/(0.323\cg^{0.125})^{4}}\right]^{1/4},
\label{eq:fitsyn}
\ee
where the pre-factor $a$ can be obtained by the normalization 
$\int_0^{0.975\chi_c} f(\cg)/\cg d\cg=1$\footnote{Although Eq.(~\ref{eq:fitsyn}) only works up to $0.95\chi_c$, we find that integrating it up to $0.975\chi_c$ gives an accurate estimation of the total energy.}. We also give a quite accurate empirical formula 
to calculate it approximately, $a=0.225 \log_{10}\left[ R\bp/{\rm (cm\cdot G)}\right]-0.67$.
Two examples are shown in Fig.~\ref{fig5}. We can see that the errors of our fittings 
are less than 4\%. For the cutoff part, we give an example of fitting for the case of 
$R\bp=10^{18}$ cm$\cdot$G by a function $0.23\exp(-6.0\times10^{24}\cg^{24.4})$
with an accuracy of 90\%. The fitting is shown in Fig. \ref{fig6}.

The above results are from the cascades induced by the particles injected at 
the same position. We here also study the case that the particles are evenly injected from 
the whole cascade region. If fresh particles are injected at a distance $l$ from 
the initial injection position, then the cascade scale of these new 
particles is $R-l$. The spectrum is then contributed by the cascades initiated by the 
particles injected from $l=0$ to $l=R$. If the injection rate of the particle source is $N_{\rm s}(l)$ 
in units of cm$^{-1}$, the normalized space-integrated spectrum can then be obtained as 
$\ft=\int_0^{R}N_{\rm s}f(\cg)dl/\int_0^R N_{\rm s}dl$. Assuming that 
$N_{\rm s}=C$ is a constant, this spectrum is $\ft=\int_0^{R}f(\cg)dl/R$. 
We draw particular attention to the cutoff part. The cutoff of the SED produced by 
the particles injected at $l$ happens at 
around $0.95\cc(l)=2.53/[2.30\log_{10} (R-l)\bp/{\rm (cm\cdot G)}-13.74]$.
For $l_1<l_2$, we have $\cc(l_1)<\cc(l_2)$, therefore the particles injected at $l<l_2$ won't contribute 
to the high-energy spectral part with $\cg>\cc(l_2)$. We assume the SED is truncated at $\sim 0.95\cc(l)$, and then the SED at the cutoff can be 
described as $f(0.95\cc(l))=0.323 (0.95\cc(l))^{0.125}$, as a result, the space-integrated spectral cutoff can be written as
\ba
\nonumber
\ft(0.95\cc(l))=\int_l^{R}f(0.95\cc)dl/R \\
\approx 0.323[0.95\cc(l)]^{0.125}(R-l)/R.\label{eq:totcut}
\ea
We show the cutoff of the cascade spectrum at $R\bp=10^{18}$ cm, the cutoff of the space-integrated spectrum (integrated from $l$=0 cm to $l=10^{18}/\bp$ cm), and a sketch of the pure 
classical synchrotron spectral cutoff (exponential-type cutoff) 
in Fig.~\ref{fig6}. The cutoff in cascades are much 
sharper than that in the pure synchrotron radiation.

\section{comparison with the cascades in black-body photon fields}
The EM cascade can also develop in a photon field, via the $\gamma\gamma$ annihilation (or pair production) and inverse Compton process. Such a process has been widely studied and applied in various environments \cite[e.g.][]{Berezinsky1975,*Agaronyan1984, *Akharonian1985akv, *Akharonian1985b,*Zdziarski1988, *Coppi1997, Aharonian2003, *Kachelriess2012, *Murase2012, *Berezinsky2016, *Wang2017}. It has been pointed out that the energy spectrum of secondary particles produced in the cascade developed in the magnetic field are similar to the spectrum of the cascade in the radiation field with a black body distribution\cite{Aharonian2003}. In this section, we will show the spectrum evolution in the later process can be fitted by a similar function to that used in the previous section.

The EM cascade can happen in photon fields when the condition $\Upsilon\equiv\epsilon \omega_0/m_e^2c^4\gg 1$ is met, where $\omega_0$ is the energy of the background photon. We here study this cascade in an isotropic black-body radiation field as an example. In the black-body radiation field, we take $\omega_0\approx2.7kT$, which is the average photon energy with a temperature $T$. Given a scale $R$, the interaction probabilities is 
\ba
\nonumber
P(\Upsilon)&=&R\int_{0}^{\infty} n(T,x)dx\int_{-1}^{1} (1-\mu)\sigma[2\Upsilon x(1-\mu)]d\mu\\
\nonumber
&=&Rn_0\int\int g(T,x)(1-\mu)\sigma[2\Upsilon x(1-\mu)]dxd \mu\\
&=&Rn_0\sigma_{\rm tot}(\Upsilon),
\ea 
where $\mu$ is the cosine of the angle between two colliding particles, $R$ is the cascade scale, $n(T,x)=n_0 g(T,x)$ is the background photon number density with $x$ being the photon energy in unit of $\omega_0$, $n_0$ is the normalized number density and $\int g(T,x)dx=1$. $\sigma$ is the differential cross section for the Compton scattering or the pair production while $\sigma_{\rm tot}(\Upsilon)$ is the weighted cross section depending only on $\Upsilon$.
We also show the normalized total interaction probabilities of these two processes $P/P_{\rm c}=\sigma_{\rm tot}/\sigma_T$ in Fig. \ref{fig1}, where $\sigma_T$ is the Thomson scattering cross section. It can be found that the interaction probabilities in photon fields are 
similar to these in magnetic fields.

\begin{figure}[]
\centering
\includegraphics[width=\columnwidth]{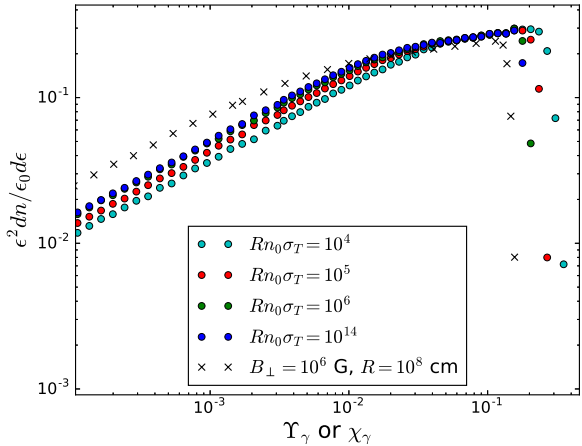}
\caption{The SEDs for different $Rn_0\sigma_T$ are presented with 
dotted curves, while an example of the spectrum in synchrotron-pair cascade is 
shown with the black cross line. \label{fig7} } \end{figure}

It has been pointed out that the resulting spectrum from the cascades in the black-body photon fields are quite similar to these in magnetic fields in the first few hundreds propagation lengths \citep{Aharonian2003}. We here study the saturated cascade using the same Monte Carlo method, and then compare it with the saturated cascade in the magnetic fields. The results are shown in Fig. \ref{fig7}. We show the spectra evolution with $Rn_0\sigma_T$. It is shown that the SED is almost unchanged when $Rn_0\sigma_T>10^6$. We then also find an 
analytical function to fit the photon SED in the range $\ug\lesssim \um$,
\be
\epsilon^2 dn/\epsilon_0 d\epsilon=[(b\ug^{0.5})^{-8}+
(0.37\ug^{0.14})^{-8}]^{-1/8},
\ee
where $\um$ is the cutoff of the spectrum. Empirically, 
we find $\um\approx0.52-0.055\log_{10}(Rn_0\sigma_T)$ and 
$b\approx0.21\log_{10}(Rn_0\sigma_T)+0.32$ for $Rn_0\sigma_T\leqslant10^6$. When 
$10^6<Rn_0\sigma_T\leqslant10^{14}$, we obtain 
$b\approx1.58$ and $\um\approx0.19$.
The accuracy is around 95\% for these empirical fittings.
Compared with Eq.~(\ref{eq:fitsyn}), we find that the spectra from cascades in
magnetic fields and photon fields are quite similar, which can also be seen in Fig. \ref{fig7}. 

\section{Summary and discussion}
In this work, we find a simple yet rather accurate analytical expression for the one-dimensional development of an EM cascade in the intense magnetic field. 
We employed the Monte Carlo method to simulate the cascade development in various combinations of parameters. We found that the normalized SED depends only on product of $R\bp$, while the species and energy of the primary particle (photon or electron) do not have influence. The SED can be described as $f(\cp)=\left[{1/(a\cg^{0.5})^{4}+1/(0.323\cg^{0.125})^{4}}\right]^{-1/4},
~{\rm when}~\cg\lesssim0.95\cc$ with $a=0.225 \log_{10}\left[ R\bp/{\rm (cm\cdot G)}\right]-0.67$. The error of the analytical expression is smaller than 4\%. We can see the photon energy spectra behave as 
$\epsilon^2 dn/\epsilon_0 d\epsilon\propto\cg^{0.5}$ at $\cg\lesssim0.003$, 
$\propto\cg^{0.125}$ at $0.003\lesssim\cg\lesssim0.1$, and a sharp cutoff 
at $\cg\sim 0.1$ (much sharper than that of the pure 
synchrotron radiation). Such a behavior is very similar to that in the EM cascade in a photon field with a black-body distribution. We found that the energy spectrum in the latter case can be fitted by a similar analytical function.
We note that such a similarity only establishes when the cascades are saturated. Otherwise, the cascade spectrum will be affected by the species (electron or photon) and the energy of the primary particle, and also the background field (magnetic field or photon field). Given that the interaction probabilities in the photon fields are scaled as 
$P\propto\sim\ug^{-1}\log(1+2\ug)$ when $\ug\gg1$, while the interaction probabilities 
in the magnetic fields are scaled as $P\propto\cg^{-1/3}$ (as shown in Fig. \ref{fig1}), the cascade spectra in magnetic field and in photon field may have big difference for a given cascade scale when the energy of the primary particle is very high, since the particle could have more interactions in the magnetic field than in the photon field.

The cascades in magnetic fields could be developed in quite different astrophysical environments and on different scales. The condition of  effective development of the cascade $\chi_0 >0.1$, combined with the range of the characteristic magnetic field in astrophysical sources, tells us at which energies the cascade could be triggered, $E >0.1 \bc/ \bp m_ec^2$. For example,  in pulsars with $B\sim10^{12}$ G, the cascade can take place at GeV energies of electrons and gamma rays; in accretion disks around solar-mass black holes with $B\sim 10^6$ G and in gamma-ray burst fireballs possibly reaching up to $B\sim 10^3$-$10^6$ G, the cascade will take place at $>$TeV energies; in the inner jets of blazars with $B\sim 1$-$100$ G,  the cascade will be triggered at energies of electrons and gamma rays $10^{17}$-$10^{19}$ eV.

\begin{acknowledgments}
We thank an anonymous referee for useful comments. We thank Dr. Kai Wang 
for helpful discussions. This work is supported by the National Basic Research Program
(``973'' Program) of China (Grant No. 2014CB845800), the National Key Research and Development
Program of China (Grant No. 2017YFA0402600), and the National Natural Science Foundation
of China (Grant No. 11573014). J.S.W. also is partially supported by the CSC scholarship.
\end{acknowledgments}

\bibliographystyle{apsrev4-1}
\bibliography{ref}

\end{document}